\documentclass{article}
\usepackage[accepted]{vietnam} 
\usepackage{natbib}
\usepackage{graphicx}

\usepackage{hyperref}
\usepackage{epstopdf}
\usepackage{amsmath}
\usepackage{txfonts}
\usepackage{color}
\usepackage{listings} 
\usepackage{times}
\usepackage{multirow}
\usepackage{float}
\usepackage{sidecap}
\usepackage{subfigure}

 \newcommand{\planck}{\textit{Planck}}  
\newcommand{\herschel}{\textit{Herschel}}
\newcommand{\NH}{N_{\rm H}} % NH
\def\NHUNIT{\ifmmode {\rm \,cm^{-2}} \else $\rm \,cm^{-2}$ \fi} % NH units
\def\nhh{\ifmmode N_{\rm H_{2}}\else $N_{\rm H_{2}}$\fi} 
\def\nh{\ifmmode N_{\rm H}\else $N_{\rm H}$\fi}
\newdimen\sa  \newdimen\sb
\def\parcs{\sa=.07em \sb=.03em
     \ifmmode \hbox{\rlap{.}}^{\scriptstyle\prime\kern -\sb\prime}\hbox{\kern -\sa}
     \else \rlap{.}$^{\scriptstyle\prime\kern -\sb\prime}$\kern -\sa\fi}
\def\parcm{\sa=.08em \sb=.03em
     \ifmmode \hbox{\rlap{.}\kern\sa}^{\scriptstyle\prime}\hbox{\kern-\sb}
     \else \rlap{.}\kern\sa$^{\scriptstyle\prime}$\kern-\sb\fi}
\begin{document}
\twocolumn[
\title{Properties of interstellar filaments as derived from  $Herschel$, $Planck$, and molecular line observations}
\titlerunning{Properties of interstellar filaments as derived from  $Herschel$, $Planck$, and molecular line observations}
\author{Doris Arzoumanian}{doris.arzoumanian@nagoya-u.jp}
\address{Laboratoire AIM, CEA/DSM--CNRS--Universit\'e Paris Diderot, IRFU/Service d'Astrophysique,\\ C.E.A. Saclay, Orme des Merisiers, 91191 Gif-sur-Yvette, France\\
Present address: Department of Physics, Graduate School of Science, Nagoya University,\\ Furo-cho, Chikusa-ku, Nagoya 464-8602, Japan}
%if there are more authors, just add more. Otherwise, please comment it.
%\author{Quang}{email@yourdomain.edu}
%\address{Faculty of Sciences, Kyushu University, Fukuoka 819-0395,Japan}

\keywords{star formation}
\vskip 0.5cm 
]

\begin{abstract}
Recent \herschel\ and \planck\ observations of submillimeter dust emission revealed the omnipresence of filamentary structures in the interstellar medium (ISM). 
The ubiquity of filaments in quiescent clouds as well as in star-forming regions indicates that the formation of filamentary structures is a natural product of the physics at play in the magnatized turbulent cold ISM. 
	An analysis of more than 270 filaments  observed with {\it Herschel}   in  8 regions   of the  Gould Belt,   shows that interstellar filaments are characterized by a narrow distribution of central width, while they span a wide column density range. 
	 Molecular line  observations of a sample of these filaments show evidence of  an increase in the  velocity dispersion of dense  filaments with column density, suggesting an evolution  in mass per unit length due to accretion of  surrounding material onto these star-forming filaments. 
	 The analyses of  \planck\  dust polarization observations  show that the mean  magnetic field along the filaments is different from that of their surrounding clouds. This points to a coupling between the matter and the $\vec{B}$-field in the filament formation process.
These observational results, derived from dust and gas tracers in total and polarized intensity, 
set strong constraints on theoretical models for  filament  formation and evolution. They also provide  important hints on the initial conditions of the star formation process from the fragmentation of dense (supercritical) filaments. 
Higher resolution dust polarization observations and large scale molecular line mapping are nevertheless required to investigate in more details the internal structure of interstellar filaments. 
\end{abstract}

\section{Omnipresence of  filamentary structures in the interstellar medium}

\begin{figure*}
		       \hspace{-0.5cm}
    \begin{minipage}{1\linewidth}
    \centering   
    %\hspace{-0.9cm}
       \resizebox{18.cm}{!}{\includegraphics[angle=0]{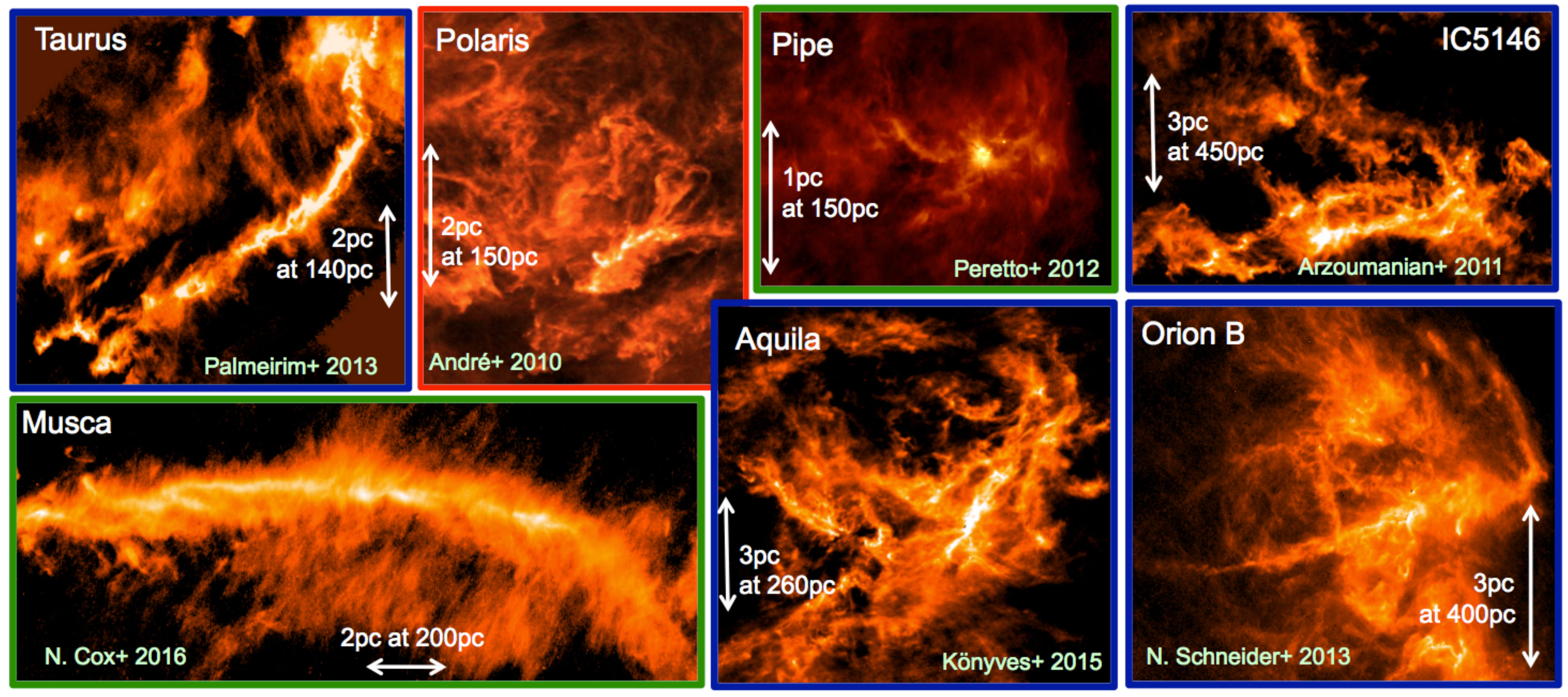}}
       \end{minipage} 
       %    \vspace{-0.4cm}
	          \caption{Column density, $N_{\rm H_2}$ [cm$^{-2}$], maps derived from  \herschel\  five-wavelength images [from 70 to 500\,$\mu$m] observed as part of  the \herschel\  Gould Belt survey  \citep{Andre2010}. 
	          These seven molecular clouds are framed with colours reflecting their star forming activity:
	          from actively star forming (blue) to mostly quiescent (red). %\citep{Peretto2012,Schneider2013,Koch2015,Men'shchikov2010,Whittet2008}  
	           }
	   \label{Herschel_Mosaic}%       
	    \end{figure*}

While molecular clouds were already known to exhibit filamentary structures   \citep[e.g.,][]{Schneider1979,Abergel1994}, 
  the omnipresence of  filaments  in the interstellar medium (ISM) and molecular clouds
 has only recently been revealed  thanks  to the high resolution and  the high dynamic range of  \herschel\  observations of the  submillimeter (submm) dust emission \citep[e.g.,][and Fig.\,\ref{Herschel_Mosaic}]{Andre2010,Molinari2010}.   
Furthermore, the all-sky maps of dust submm emission observed by \planck\ in total intensity, as well as in polarized intensity, 
emphasize the large-scale, hierarchical filamentary texture  of the Galactic ISM \citep{planck2015-XIX,planck2016-XXXII}.
 \herschel\ and \planck\ data show that interstellar matter is organized in   web-like networks of filaments, 
which appear to be formed as a natural result of the physics at play in the ISM.
The presence of  interstellar filaments  in both  star-forming regions as well as in  non-star-forming quiescent clouds, alludes to a formation process of filaments  preceding any star-forming activity \citep{Andre2010}. The spatial distribution of prestellar cores and protostars extracted from \herschel\ images observed mainly along the densest filaments  \citep{Konyves2015} indicates  that the properties of interstellar filaments may be a key element defining  the initial conditions required for the onset of star formation  \citep{Andre2014}.

 \begin{figure*}[ht]
		        \hspace{-0.1cm}
    \begin{minipage}{1\linewidth}
    \centering   
       \resizebox{17.cm}{!}{\includegraphics[angle=0]{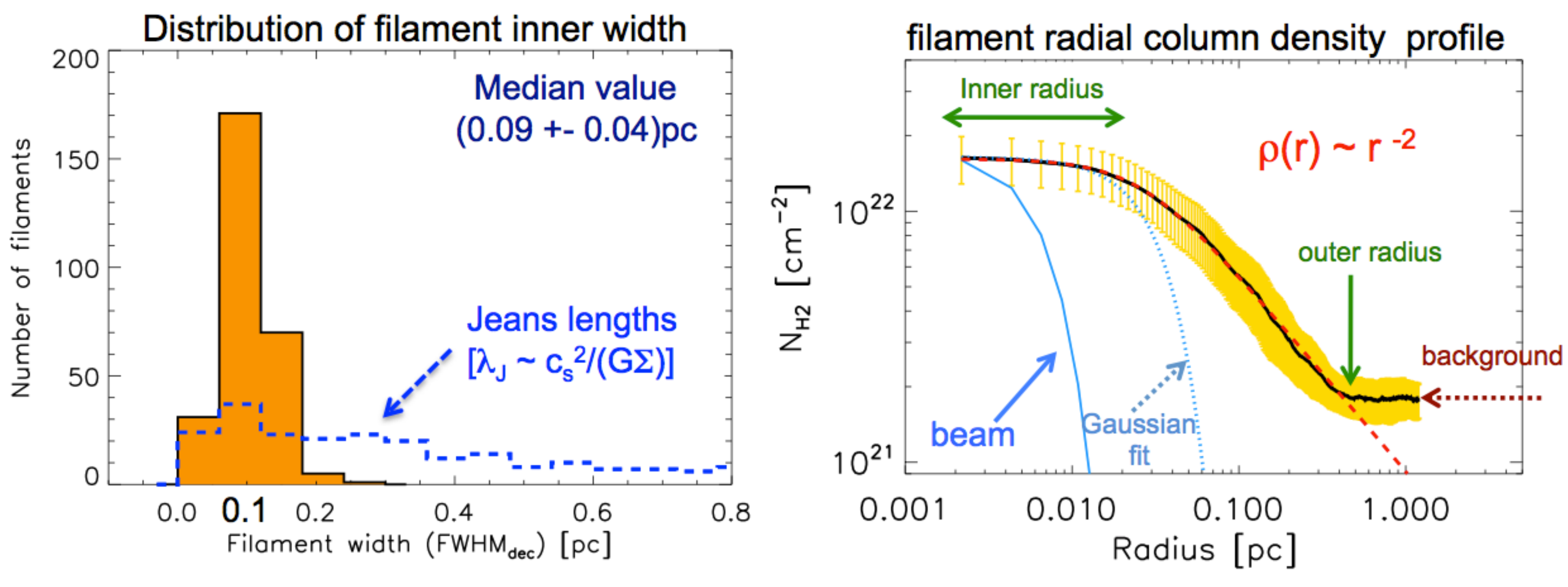}}
       \end{minipage}   
     %  \vspace{-0.5cm}
	          \caption{
	       {\bf Left:}     Distribution of deconvolved FWHM widths for  278 filaments observed in 8 different regions (black solid histogram, filled in orange), with a
	median value of  $0.09\pm0.04$\,pc.
	For comparison, the blue dashed histogram represents the distribution of central (thermal) Jeans lengths 	of the filaments. 
	{\bf Right:}  Radial column density profile averaged along the length of the B211/13 filament in Taurus \citep{Palmeirim2013}.	 The median absolute deviation of the radial profiles along  the filament length is shown in yellow. The profile is well fitted with a {\it Plummer-like} function (red dashed curve) where the density decreases as $r^{-2}$ at large radii \citep[][]{Arzoumanian2011,Palmeirim2013}. 
%\citep{Peretto2012,Schneider2013,Koch2015,Men'shchikov2010,Molinari2010,Whittet2008}   
} 
	   \label{ProfWidth}%       
	    \end{figure*}
	    
%\citep{Peretto2012,} 
% \citep{Schneider2013} 
% \citep{Goldsmith2008} 
% \citep{Heitsch2013}
% \citep{Koch2015}
%\citep{Men'shchikov2010}
%\citep{Molinari2010}
%\citep{Sousbie2011}
%\citep{Whittet2008}
 
Hence, characterizing  the observed filament properties in detail, combining  tracers of gas and dust in total and polarized intensities  is essential  to make progress in our understanding of the physical processes involved in the formation and evolution of interstellar filaments and their role in the star formation process. 

 In the following, I discuss the main results on the properties of interstellar filaments derived from \herschel, \planck, and molecular line observations. 
 \herschel\ continuum observations are essential to describe the filament (column) density distribution. These data are complemented with    ground based molecular line observations to access to the kinematics of the filamentary structures, while \planck\ dust polarization observations give unprecedented information on the structure of the magnetic field and its connection with interstellar matter. 
 These results are presented in the context of a new paradigm of star formation, which is closely linked to the formation and fragmentation of self-gravitating filaments.

\section{Filament properties as derived from \herschel\ observations of nearby clouds}\label{profiles}
	    
 Statistical analysis of nearby  interstellar filaments  has been possible thanks to the \herschel\ Gould Belt survey observations  \citep{Andre2010}, which are ideal to characterize the filament properties, providing the resolution, the sensitivity, and the statistics needed for such studies.  \herschel\ observations of a large number of clouds with different star formation activities  and environments have been analysed \citep[e.g.,][and Fig.\,\ref{Herschel_Mosaic}]{Men'shchikov2010,Peretto2012,Schneider2013}.

Detailed measurements of the radial column density profiles (see Fig.\,\ref{ProfWidth}-Right) derived from \herschel\ column density maps show that interstellar filaments are characterized by a narrow distribution of central widths, around 0.1 pc (Fig.\,\ref{ProfWidth}-Left), while they span almost three orders of magnitude in central column density as can be seen on the left hand side of Fig.\,\ref{VelDisp_coldens}\,\citep[][and Arzoumanian et al., in prep.]{Arzoumanian2011}.

This characteristic filament width \citep[cf.,][for an independent analysis]{Koch2015}  is well resolved by \herschel\ observations of the Gould Belt clouds as can be seen on the radial profile shown in Fig.\,\ref{ProfWidth}-Right and Fig.\,\ref{VelDisp_coldens}-Left, where  the measurements of the filament widths lie above the horizontal lines corresponding to the resolution limits. 
 This typical filament width of 0.1\,pc is also in contrast with the much broader distribution of central Jeans lengths, $\lambda_{\rm J} \propto c_{\rm s}^2/(GN_{\rm H_{2}}^0)$, 
	from 0.02 to 1.3\,pc (for $T=10$\,K),  implying that these filaments are not in hydrostatic equilibrium. 
	
	These filaments also span a wide range in mass per unit length  ($M_{\rm line}$),  estimated from their column density profiles. 
	The mass per unit length of a filament,  is a very important parameter, which defines its ``stability":  a filament is  subcritical, unbound when its  mass per unit length is smaller than the critical value $M_{\rm line, crit} = 2\, c_s^2/G \sim 16\, M_\odot$/pc, where  $c_{\rm s} \sim 0.2$~km/s is the isothermal sound speed for $T \sim 10$~K, and $G$ is the gravitational constant \citep[][]{Ostriker1964}. A filament is supercritical, unstable for radial collapse and fragmentation, when $M_{\rm line}>M_{\rm line, crit}$ \citep{Inutsuka1997}. 

\section{Internal velocity dispersions of interstellar filaments derived from molecular line observations}

\begin{figure*}
	    	\hspace{-0.05cm}
	    \begin{minipage}{1\linewidth}
    		\centering
       		\resizebox{8.cm}{!}{\includegraphics[angle=0]{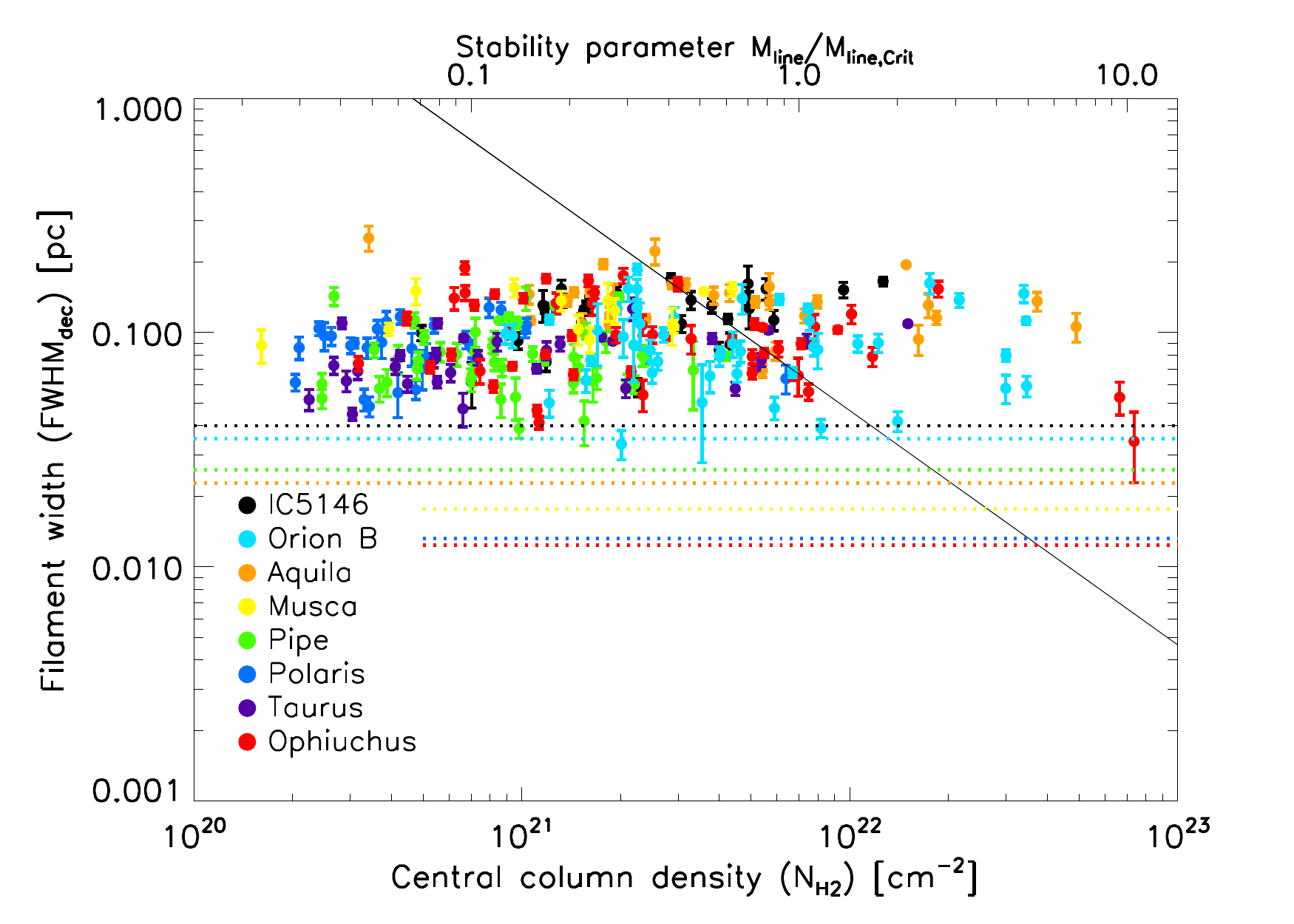}}
		\hspace{0.05cm}
       		\resizebox{8.cm}{!}{\includegraphics[angle=0]{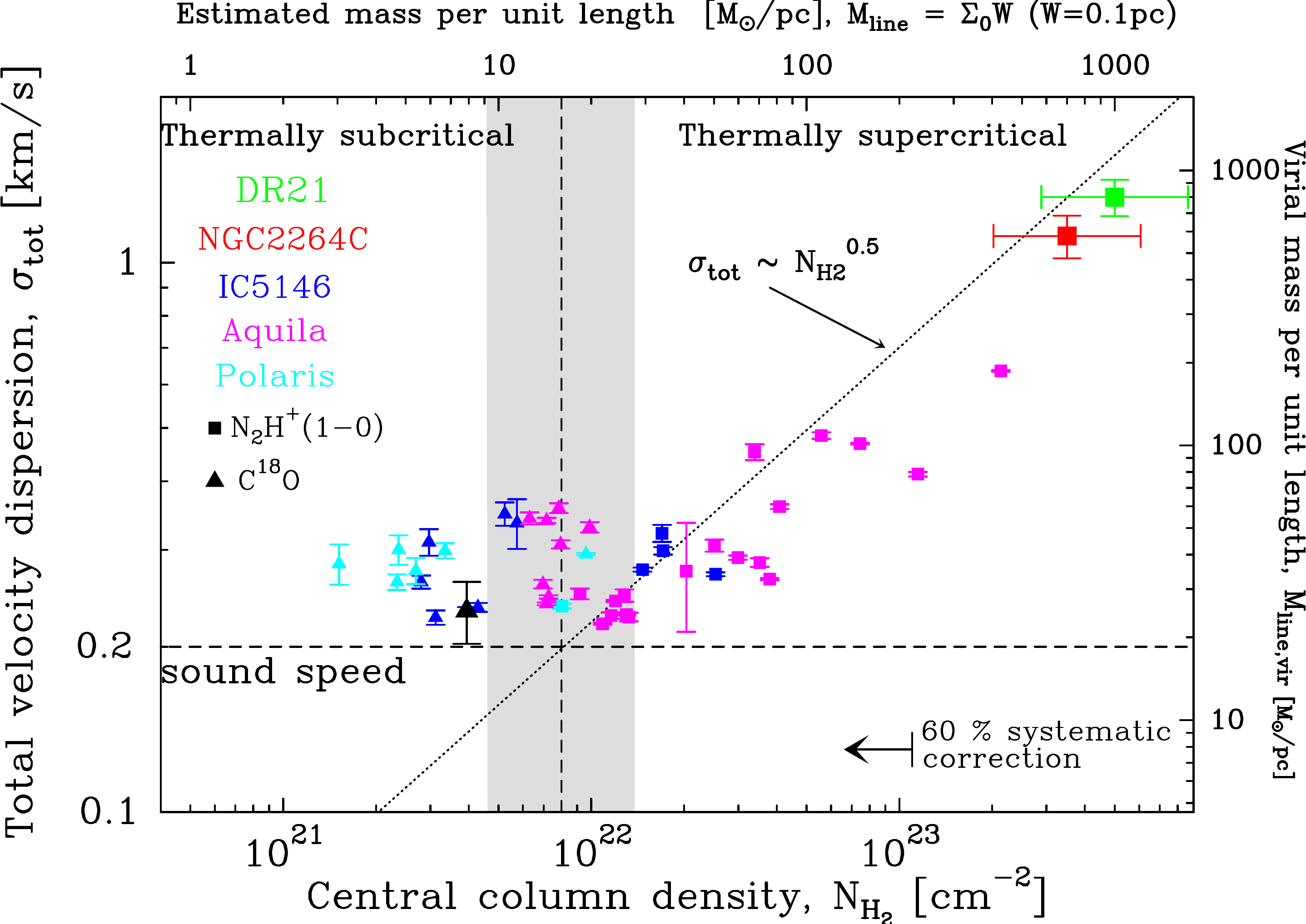}}
		\hspace{0.05cm}
       	\end{minipage}   
	  	%\vspace{-0.8cm}
	         \caption{
	         {\bf Left:}
	         Mean deconvolved width versus  background subtracted central column density for the filament sample analysed in 8 regions (indicated on the plot). The spatial resolutions of the column density maps  
	 are marked by the horizontal dotted lines. The solid line running from top left to bottom right shows the central  Jeans length as a function of central column density. The  upper $x$-axis scale is an estimate of the filament  mass per unit length in units of the thermal critical value $M_{\rm line,crit} = 2c^2_{\rm s} /G$, where $M_{\rm line}\propto W N_{\rm H_{2}}^0$  with 	 $W = 0.1$~pc  \citep[][]{Arzoumanian2011}.
	           {\bf Right:}
		Filament total velocity dispersion versus observed central column density. 	
				The vertical dashed line marks the boundary between thermally subcritical and thermally supercritical filaments where the estimated mass per unit length $M_{\rm line}$ is approximately equal to the critical value $M_{\rm line,crit} \sim$ 16~M$_{\odot}$/pc for T=10~K, equivalent to a column density of $8\times10^{21}$~cm$^{-2}$. The grey band shows a dispersion of a factor of 3 around this nominal value. The dotted line running from the bottom left to the top right corresponds to $ \sigma_{\rm tot} \propto {N_{\rm H_2}}^{0.5} $ \citep[][]{Arzoumanian2013}. 	
	 } 
   \label{VelDisp_coldens}%     
    \end{figure*}

The total velocity dispersion ($\sigma_{\rm tot}$)  of selected positions towards a sample of  46 filaments, derived from ($^{13}$CO, C$^{18}$O, and N$_2$H$^+$) 
molecular line observations, are presented in Fig.\,\ref{VelDisp_coldens}--Right. Thermally subcritical and nearly critical filaments have 
transonic velocity dispersions ($c_{\rm s} \lesssim \sigma_{\rm tot} < 2c_{\rm s}$) independent of column density and are gravitationally unbound.
The velocity dispersion of thermally supercritical filaments increases as a function of their column density  (roughly as $ \sigma_{\rm tot} \propto {N_{\rm H_2}}^{0.5} $). 
  These  measurements confirm that there is a critical threshold in $M_{\rm line}$ above which  filaments are self-gravitating and below which they are unbound. The position of this threshold, is consistent within a factor of two with the critical value  $M_{\rm line,crit} \sim$ 16~M$_{\odot}$/pc for T=10~K, equivalent to a column density of $8\times10^{21}$~cm$^{-2}$  \citep[][]{Arzoumanian2013}. 
 
  These observations show that  the mass per unit length  of supercritical filaments is close  to their virial mass per unit length $M_{\rm line,vir}=2\sigma_{\rm tot}^2/G$ \citep{Fiege2000} where  $\sigma_{\rm tot}$ is the observed total velocity dispersion (instead of the thermal sound speed used in the expression of $M_{\rm line,crit}$).

	    We suggest that the large velocity dispersions of supercritical filaments is not a result of the supersonic interstellar turbulence but may be driven by gravitational contraction/accretion \citep{Arzoumanian2013}. 
Mass accretion is indirectly suggested by  transverse velocity gradients observed across a self-gravitating filament in Taurus.  
The systemic velocities \citep{Goldsmith2008} of the observed emission in the north and south sides of the filament are 
  redshifted and blueshifted, respectively, with respect to the velocity of the emission observed towards the B211/13 filament \citep{Palmeirim2013}. 
Such  a velocity field pattern 
 may indicate convergence of matter  onto the densest parts of the supercritical filament.
This is also compatible with theoretical models for the evolution of supercritical filaments \citep{HennebelleAndre2013,Heitsch2013}. Such models  put forward the role of  continuous accretion, which may be a  physical reason to explain  the observed  properties of  supercritical filaments.

\section{Magnetic field structure as derived from \planck\ dust polarization observations}

\begin{figure*}
   %   \hspace{2.cm}
  \centering
      \includegraphics[width=0.99\textwidth]{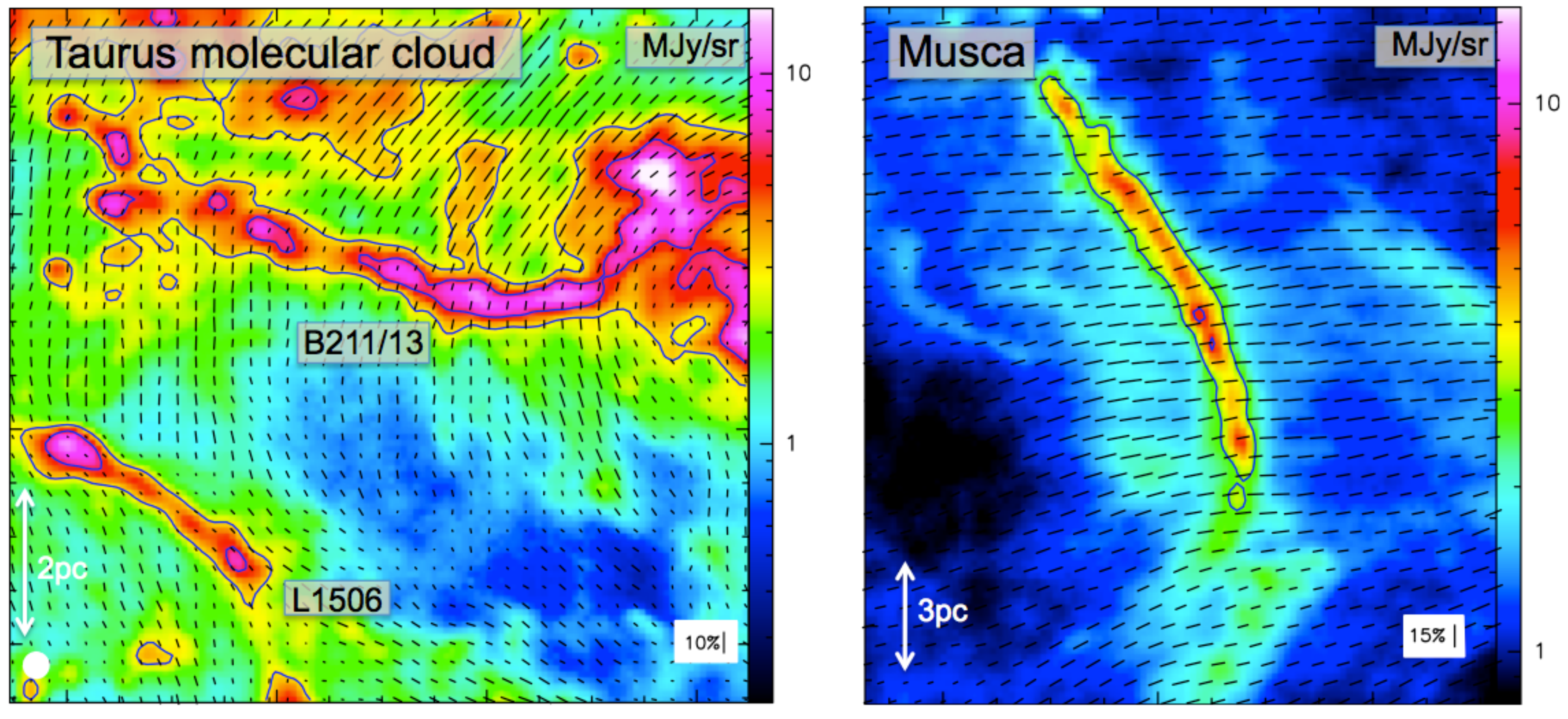}
      %\vspace{-0.5cm}
	          \caption{
	          \planck\ 353\,GHz (850\,$\mu$m) total dust intensity (Stokes $I$) maps at a resolution of 4\parcm8, towards the Taurus B211/13 and L1506 filaments  {\bf (Left)} and  the Musca filament    {\bf (Right)}. The maps are in Galactic coordinate system.	 The blue contours show the  levels of 3 and 6 MJy\,sr$^{-1}$.          
	           The black segments show the $\vec{B}_{\rm POS}$-field orientation ($\psi$+90$^{\circ}$). The length of the pseudo-vectors is proportional to the polarization fraction. The polarization angles and fractions are computed at a resolution of 9\parcm6 (indicated by the white filled circles on the left hand side map)  for increased S/N \citep{planck2016-XXXIII}.  	   
	              } 	  
   \label{planckMaps}
   \end{figure*}

Dust polarization observations are  essential  to infer the orientation of the magnetic field ($\vec{B}$)  component projected on the plane of the sky (POS). 
While the observed polarization fraction ($p$) depends on several parameters \citep[dust polarization properties, grain alignment efficiency, and $\vec{B}$-field structure, e.g.,][]{Hildebrand1983}, the observed  
polarization angle ($\psi$) derived from dust polarized emission 
is perpendicular to orientation of the  $\vec{B}$-field component on the POS ($\vec{B}_{\rm POS}$)  averaged along the line of sight (LOS). 
\planck\  observations  at 353\,GHz provide  the first fully sampled maps of the polarized dust emission towards interstellar filaments and their backgrounds, providing  unprecedented insight into  the $\vec{B}$-field structure. 
The first striking result is the impressively ordered  structure of $\vec{B}_{\rm POS}$ from the largest Galactic scales down to the smallest scales probed by  \planck\ observations, $\sim$0.2\,pc in nearby molecular clouds  \citep[Fig.\,\ref{planckMaps}, and][]{planck2015-XIX,planck2016-XXXV}.
   
    To quantify the polarized intensity  observed towards the filaments,  
we derive  radial profiles perpendicular to their crests and 
averaged  along their length to increase the signal-to-noise ratio in polarization, while keeping the highest resolution (4\parcm8) of the \planck\ data.
We describe the observations as a two-component model   and separate the emission of the filaments from  that of their background (surrounding cloud). 
  This allows us to characterize and  compare the polarization properties of each emission component: the filament and its background.
This is an essential step in measuring  the intrinsic polarization fraction ($p$)  and polarization angle ($\psi$) of each emission component. 
The analyses show that both the polarization angle and fraction measured at the intensity peak of the filaments (before background subtraction) differ from their intrinsic values (after background subtraction) as described in \citep{planck2016-XXXIII}. 

\begin{figure*}
  \centering
      \includegraphics[width=0.99\textwidth]{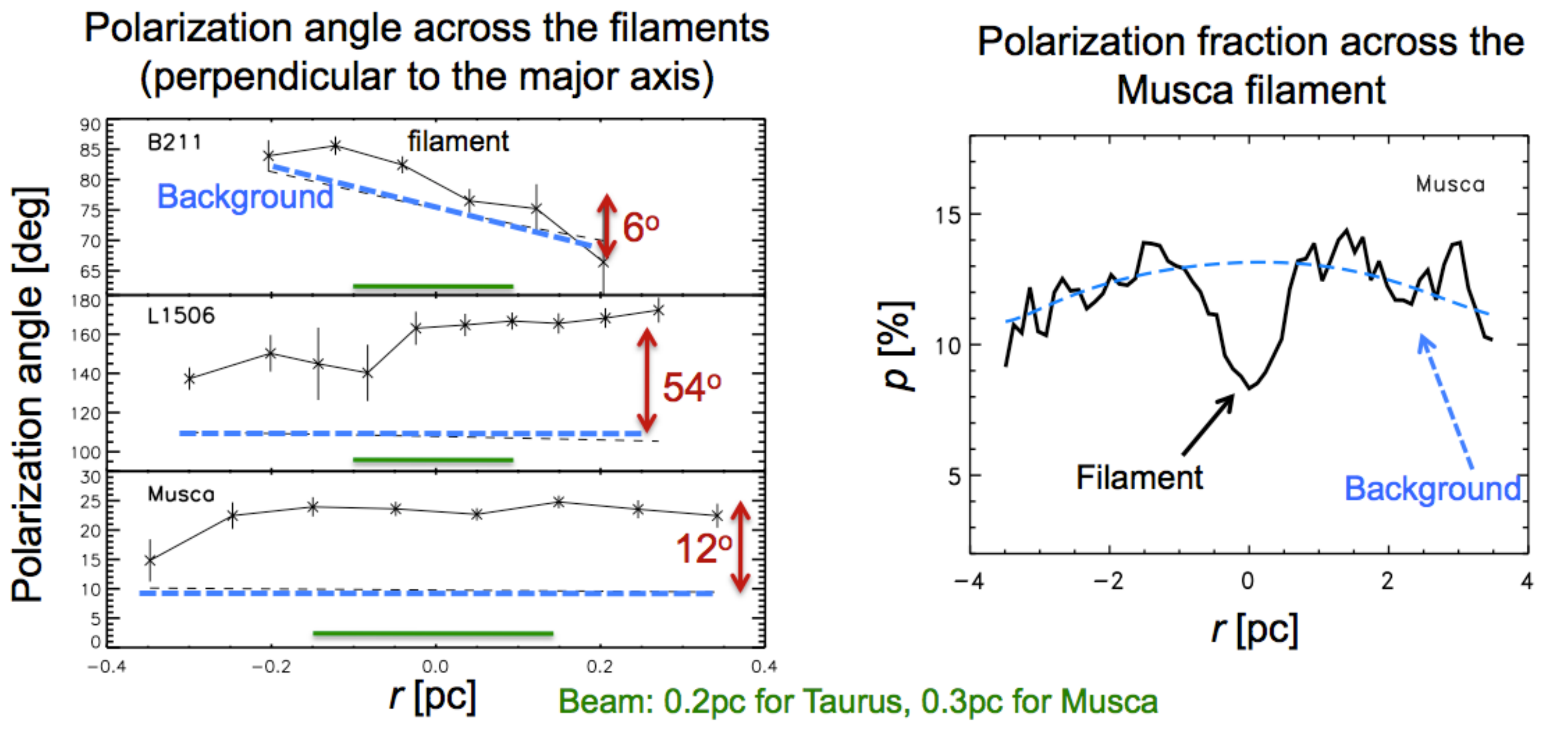}
    %  \vspace{-0.5cm}
	          \caption{{\bf Left:}          
	          Filament intrinsic polarization angle (tracing the angle of $\vec{B}_{\rm POS}$) across the crests of the  B211, L1506, and Musca filaments.  The $x$-axis shows the radial distance from the filament crest. 
   The crosses are  data points computed from $Q$ and $U$ background subtracted maps. 
The dashed line represents the background polarization angle.
The difference between the polarization angle of the filament and that of the background is indicated on the plots \citep[taken from Table\,2 in][]{planck2016-XXXIII}. 
	   {\bf Right:}    
	   Observed  profile (in black) of the polarization fraction ($p$) perpendicular to the crest of the Musca filament.  
        The dashed blue curve shows the polarization fraction of the background. It is derived from the Stokes  $I$, $Q$, and $U$ parameters of the background.      
   } 	                
	   \label{PolarParam}       
\end{figure*}

The left panel of Fig.\,\ref{PolarParam} shows the profile of $\psi$  across the Musca, B211, and L1506 filaments. 
In all three cases,  we measure variations in the polarization angle intrinsic to  the filaments ($\psi_{\rm fil}$) with respect to that of their 
backgrounds ($\psi_{\rm bg}$).  These variations are found to be coherent along the pc-scale length of the filaments.  
The  differences between $\psi_{\rm fil}$ and  $\psi_{\rm bg}$ for two of the three filaments are larger 
than the dispersion of the polarization angles across and along the filaments  \citep[see Table\,2 in][]{planck2016-XXXIII}.  Hence, these differences are not random fluctuations and they indicate a change in the orientation 
of the POS component of the magnetic field between the filaments and their backgrounds.

 The observations  show a decrease in  the polarization fraction $p$ from the background to the  Musca filament (Fig.\,\ref{PolarParam}--right), as well as towards the Taurus B211 and L1506 filaments \citep[see Fig.\,10 of][]{planck2016-XXXIII}. 
 A decrease  in $p$ with the total column density $\NH$ (i.e., from the backgrounds to the filament crests), has been already shown in previous studies. 
This decrease has been usually interpreted as due to the turbulent component of the field and/or variations of dust alignment efficiency with increasing column density \citep[e.g.,][and references therein]{Whittet2008,Jones2015}.
In our study, the bulk of the drop in $p$ within the filaments  cannot be explained by random fluctuations of the orientation of the magnetic field  because they are too small ($\sigma_{\psi}<10^\circ$).
We argue that the observed drop in $p$ towards the filaments may be due to the 3D structure  of the magnetic field:   both its orientation in the POS and with respect to the POS \citep{planck2016-XXXIII}.
Indeed, the observed changes of $\psi$ are direct evidence of variations of the orientation of the POS projection of the magnetic field. 
The systematic variations of $\psi$  suggest changes of the angle of the magnetic field with respect to the POS. This angle must statistically  vary as much as $\psi$, contributing to the observed decrease of $p$ in the filaments. 
The observed variation of $\psi$ between the filaments and their backgrounds always depolarizes the total emission, due to the integration of  
the emission along the LOS of two emission components where the angle of $\vec{B}_{\rm POS}$ varies  \citep[see Fig.\,\ref{PolarParam} and][]{planck2016-XXXIII}.

 The inner structure of the filaments \citep[as seen, e.g., with \herschel, cf.,][]{Palmeirim2013,Cox2016},   is not resolved, but  at the smallest scales accessible with \planck\ ($\sim$0.2\,pc towards the nearby clouds), the observed changes of $\psi$ and $p$ (derived from \planck\ polarization data at  the resolution of 4\parcm8) hold some information on the magnetic field structure within filaments \citep{planck2016-XXXIII}.  They show that both the mean  field and its fluctuations in the filaments are different from those in the background clouds, which points to 
a coupling between the matter and the $\vec{B}$-field in the filament formation process.

\section{Summary and conclusions}

 The ubiquity of filaments in both quiescent clouds and active star-forming regions, where they are associated with the presence of prestellar cores and protostars \citep{Konyves2015}, supports the view 
that filaments are (first) form in the ISM 
and the densest of them  fragment into star-forming cores \citep{Andre2014}. 
The observational finding of a filament uniform 0.1\,pc width \citep{Arzoumanian2011}  sets  strong constraints on the physics at play in the ISM. This result has been recently observed (at higher-resolution)  from the ground in regions farther away than the Gould Belt \citep{Hill2012Artemis,Andre2016}.  

 Interestingly 0.1\,pc corresponds to the sonic scale below which interstellar turbulence becomes subsonic in diffuse, non-star-forming gas. This occurrence, along with the observed thermal velocity dispersion of low column density filaments,  suggests that large-scale turbulence may be  a main player in the formation of the filamentary web observed  in molecular clouds \citep{Arzoumanian2011}.
On the other hand,  the increase of the  non-thermal
  velocity dispersion of supercritical, self-gravitating filaments, with column density,   may indicate the generation of internal turbulence due to  gravitational accretion. 
  This may be an explanation for the observed constant width of self-gravitating collapsing filaments  \citep{Arzoumanian2013}.

	 While  the dissipation of interstellar turbulence provides a plausible mechanism for filament formation, the observed organization between the magnetic field lines and the intensity structures, derived from the analysis of \planck\ data, indicates that the $\vec{B}$-field  plays a dynamically important role in shaping the  interstellar matter
	 \citep{planck2016-XXXIII,planck2016-XXXV}.  In particular, they may be a key element to understand the channel of mass flows in the ISM.
	 	The fact that most prestellar cores  lie in dense, self-gravitating filaments \citep{Konyves2015} suggests that gravity is a major driver in the evolution of  supercritical  filaments and their fragmentation  in star-forming prestellar cores.
		
		The combination of these observational results, derived from dust and gas tracers in total and polarized intensity, give strong constraints on our understanding of the formation and evolution of filaments in the ISM, which provides important clues to the initial conditions of the star formation process along supercritical filaments. Higher resolution dust polarization observations and large scale molecular line mapping are nevertheless required to investigate in more details the internal velocity and magnetic field structures of interstellar filaments. 
		
%\begin{figure*}
%\vskip -0.5cm
%\centering
%$\begin{array}{cc}
%\includegraphics[angle=0,width=13.cm]{f4.eps} 
%\end{array}$
%\caption{
%Plasma beta (color)  and velocity vectors on the equatorial plane (left) and $y=0$ plane (right) in the mass accretion phase \cite{machida14b}.
%The density contour is overplotted by contour lines. 
%}
%\label{fig:4}
%\vspace{-0.5cm}
%\end{figure*}

\section*{Acknowledgments}

DA has received support from the European Research Council grant ORISTARS (No.\,291294) 
and is currently an International Research Fellow of the Japan Society for the Promotion of Science (FY2016).

%%\section*{References}
%\begin{thebibliography}
%\bibitem[Bate(1998)]{bate98} Bate, M.~R.\ 1998, ApJL, 508, L95
%\bibitem[Bate(1998)]{machida14b} Bate, M.~R.\ 1998, ApJL, 508, L95

%Or link to external ADS style bib
%\bibliographystyle{apj} 
%\bibliography{Arzoumanian_sfde16}

%\end{thebibliography}

\end{document}